\documentclass[dvipdfmx]{jpsj-suppl}
\usepackage{txfonts} 

\title{The {\it s}-Process Nucleosynthesis in Extremely Metal-Poor Stars as the Generating Mechanism of Carbon Enhanced Metal-Poor Stars}

\author{Takuma \textsc{Suda}$^{1}$, Shimako \textsc{Yamada}$^{2}$, and Masayuki Y. \textsc{Fujimoto}$^{3}$}

\inst{$^{1}$Research Center for the Early Universe, The University of Tokyo, Hongo 7-3-1, Bunkyo-ku, Tokyo 113-0033, Japan \\
$^{2}$Department of Cosmosciences, Hokkaido University, Kita 10 Nishi 8, Kita-ku, Sapporo 060-0810, Japan \\
$^{3}$Faculty of Engineering, Hokkai-Gakuen University, Minami 26 Nishi 11, Chuo-ku, Sapporo 064-0926, Japan}

\email{suda@resceu.s.u-tokyo.ac.jp}

\recdate{September 7, 2016}

\abst{
The origin of carbon-enhanced metal-poor (CEMP) stars plays a key role in characterising the formation and evolution of the first stars and the Galaxy since the extremely-poor (EMP) stars with [Fe/H] $\leq -2.5$ share the common features of carbon enhancement in their surface chemical compositions. The origin of these stars is not yet established due to the controversy of the origin of CEMP stars without the enhancement of {\it s}-process element abundances, i.e., so called CEMP-no stars.
In this paper, we elaborate the {\it s}-process nucleosynthesis in the EMP AGB stars and explore the origin of CEMP stars.
We find that the efficiency of the {\it s}-process is controlled by O rather than Fe at [Fe/H] $\lesssim -2$.
We demonstrate that the relative abundances of Sr, Ba, Pb to C are explained in terms of the wind accretion from AGB stars in binary systems.}

\kword{metal-poor star, evolution, AGB, {\it s}-process nucleosynthesis}

\begin{document}
\maketitle

\section{Introduction}

Carbon-enhanced metal-poor (CEMP) stars are commonly found among metal-poor stars having the iron abundance of [Fe/H] $\lesssim -2$.
They comprise more than 20 \% of the known metal-poor stars \cite{Frebel2015}, and are important probes for the star formation history in the early Galaxy.

CEMP stars are divided into several subgroups depending on the abundances of neutron-capture elements \cite{Beers2005}.
The most well accepted and common populations are CEMP-{\it s} and CEMP-no stars.
They are classified by the abundance of barium, which is the representative of the {\it s}-process.
Thanks to the many efforts of abundance analyses of extremely metal-poor ([Fe/H] $\lesssim -3$) stars, it is established that CEMP-no stars dominate the population at very low metallicity \cite{Aoki2007}.

The origin of CEMP-no stars is still in controversy, although there is an argument that CEMP-{\it s} and CEMP-no stars originated from different populations \cite{Starkenburg2014}, mainly regarding the binarity of CEMP populations.
However, the distribution of [Ba/C] as a function of metallicity is continuous rather than discrete between CEMP-{\it s} and CEMP-no stars (see the plot using the SAGA database \cite{Suda2008,Suda2011,Yamada2013}), which implies that CEMP-{\it s} and CEMP-no stars have the same origin.
This can be interpreted that CEMP-no stars should also have or had binary companions since it is almost established that the abundances of CEMP-{\it s} stars are the result of binary mass transfer.

In addition, CEMP-no stars show smaller carbon enhancement than CEMP-s stars, whatever the trend of carbon abundance is.
This also implies that CEMP-no stars belong to longer periond binaries if the amount of C dredged-up to the surface in AGB stars is constant.

It is expected that C and {\it s}-process elements are transported together during the binary mass transfer, because these elements are synthesized in the helium burning layers, dredged-up to the surface in a primary star, and transferred to a secondary star, keeping the abundance ratio constant.

In this study, we explore the hypothesis that all the CEMP stars experienced binary mass transfer from former AGB stars.
We compute the {\it s}-process in EMP AGB stars to provide a general picture of the origin of CEMP stars.



\section{Models}

We have considered three channels of the {\it s}-process in EMP AGB stars:
(1) the convective $^{13}$C burning which is triggered by the hydrogen mixing into the helium-flash convective zones \cite{Fujimoto2000,Suda2004,Campbell2008,Suda2010},
(2) the convective $^{22}$Ne burning which is the {\it s}-process in the helium-flash convective zones with a neutron source from $^{22}{\rm Ne} \left( \alpha, n \right) ^{25}{\rm Mg}$ \cite{Lugaro2012}, and
(3) the radiative $^{13}$C pocket which is triggered by the hydrogen mixing at the bottom of the envelope during the third dredge-up.

We use a nuclear network code based on one-zone approximation which includes more than 300 isotopes with proton-, $\alpha$-, and neutron-capture reactions and $\beta$-decays \cite{Nishimura2009}.
According to the prescription in \cite{Aikawa2001}, we followed the time evolution of the temperature of the nucleosynthesis site during the helium shell flashes.

Table~\ref{tab:model} shows the list of models computed in this study.
Each column denotes the model name, the maximum temperature at the bottom of the helium-flash convective zone, the proper pressure of the helium shell flash, and mass and radius at the bottom of the helium burning shell.
We have tried to cover as wide parameter ranges of helium shell flashes as possible to represent any events of mixing and dredge-ups in AGB progenitors.

\begin{table}[tbh]
\caption{Characteristics of helium flash models.}
\label{tab:model}
\begin{tabular}{lllll}
\hline
Model & $\log T_{p}$ & $\log P_{*}$ & $M_{c}$ & $r_{c}$ \\
      & [K]          & dyne cm$^{-2}$  & $M_{\odot}$     & $R_{\odot}$ \\
\hline
M857 & 8.57 & 20.07 & 1.30 & $5.30 \times 10^{-3}$ \\
M853 & 8.53 & 20.10  & 1.15 & $7.94 \times 10^{-3}$ \\
M850 & 8.50 & 20.10  & 1.00 & $9.88 \times 10^{-3}$ \\
M8485 & 8.485 & 20.11 & 0.93 & $1.08 \times 10^{-2}$ \\
M847 & 8.47 & 20.15 & 0.85 & $1.18 \times 10^{-2}$ \\
M844 & 8.44 & 20.34 & 0.70 & $1.41 \times 10^{-2}$ \\
\hline
\end{tabular}
\end{table}

\section{Results}

\begin{figure}[tbh]
\begin{center}
\includegraphics[width=0.5\textwidth]{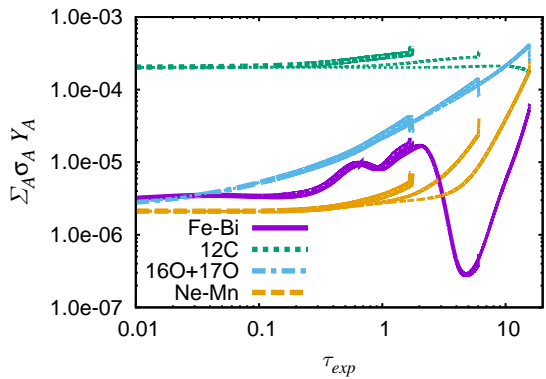}
\caption{Contribution to neutron absorption, provided by the product of the cross section and abundance of target elements,  as a function of neutron exposure for the Model M844 with [Fe/H] $= -2$.}
\label{fig:texp}
\end{center}
\end{figure}

Figure~\ref{fig:texp} shows contribution to the neutron absorption by the elements in consideration as a function of the neutron exposure which is the expected number of neutrons absorbed by the target elements in the helium flash convective zone.
The contribution to neutron absorption reactions can be measured by the product of the neutron capture cross section and the abundance.
Neutron exposures correspond to time or the amount of mixing at the maximum luminosity of the helium burning.
All the lines are taken from the result of model M844 with [Fe/H] $= -2$.
The different lines with the same line type, correspond to different mixing parameters with $\delta_{\rm mix} (= ^{13}$C / $^{12}$C) $= 0.003$, $0.03$, and $0.3$, in increasing order of the maximum of the neutron exposure.

The solid line shows the change of the neutron-capture efficiency for all the elements from Fe to Bi.
The sharp drop at $\tau \approx 3$ shows the first end of the {\it s}-process starting from the iron group elements, that is, all the initial heavy elements are converted to Bi and Pb.

After the end of the first {\it s}-process, neutron captures by light elements like Ne and Mg produce Fe, and activate the production of the {\it s}-process elements again.
The dashed lines show the neutron-absorption efficiency for Ne to Mn.
They increase sharply at $\tau \approx 3$ as that of the iron group elements drops.
The dotted lines show that $^{12}$C is the most efficient absorber of neutrons simply due to its large abundance.
But they produce neutrons again by the neutron recycling reactions \cite{Gallino1988,Nishimura2009}.
Therefore, $^{12}$C is actually not a neutron poison.
Oxygen is a neutron poison at this metallicity thanks to this reaction.
Iron group elements share comparable contributions to the {\it s}-process only at the beginning of the nucleosynthesis.
Neutrons produced by the reaction, $^{13}{\rm C} \left( \alpha, n \right) ^{16}$O, are mainly absorbed by $^{16}$O and $^{17}$O rather than by the iron group elements when $\tau > 0.1$, as shown by the dot-dashed lines.

Our analysis shows that the {\it s}-process is mainly controlled by O and not by Fe.
In our simulations, the final abundance distribution of the {\it s}-process elements is independent of metallicity at [Fe/H] $\lesssim -2$.
It is generally thought that neutron per seed nuclei increases with decreasing metallicity at [Fe/H] $\geq -2.5$, which leads to the more efficient production of heavy-{\it s} elements \cite{Karakas2014}.
However, it is not true at very low metallicity.

\begin{figure}[tbh]
\begin{center}
\includegraphics[width=1.0\textwidth]{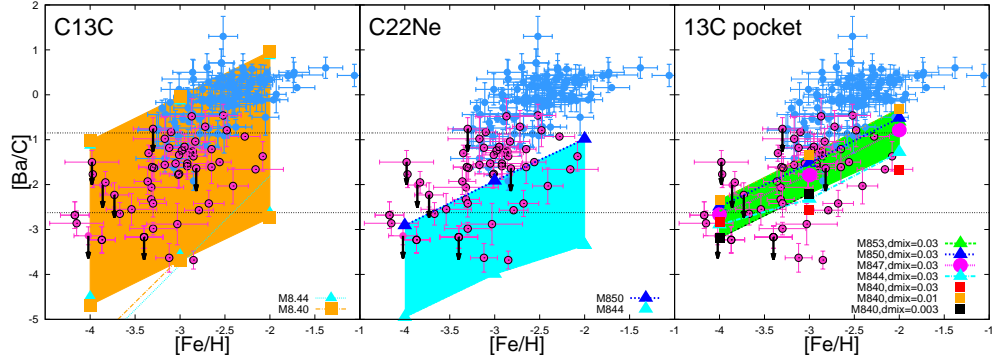}
\caption{Comparison of models with the observations of CEMP stars. The panel on the left, in the center, and on the right is the hydrogen mixing in metal-poor AGB stars, neutron source by the $\alpha$-capture of $^{22}$Ne, and the radiative $^{13}$C pockect, respectively. Blue and magenta circles denote the observed CEMP-{\it s} and CEMP-no stars taken from the SAGA database, respectively.}
\label{fig:obs}
\end{center}
\end{figure}

Figure~\ref{fig:obs} shows the comparison of our models with the observations of CEMP stars taken from the SAGA database.
The colored areas shows the final abundances covered by changing the model parameters.
It is clear that the convective $^{13}$C burning in metal-poor AGB stars (left panel) cover all the observed values, irrespective of CEMP-{\it s} or CEMP-no.
The coverage of $^{22}$Ne source (center) is smaller than the convective $^{13}$C burning, but this process partly covers the domain of CEMP-no stars.
The radiative $^{13}$C pocket (right) plays the least role among the three modes of the {\it s}-process.
In particular, the size of the $^{13}$C pocket is, in general, too small to produce {\it s}-process elements sufficient to explain the observed values of [Ba/C].

\section{Conclusions}

We have explored the possiblity of the binary scenario to reproduce the observed abundances of carbon-enhanced metal-poor stars with and without the enhancement of {\it s}-process elmements (CEMP-{\it s} and CEMP-no, respectively), using nucleosynthesis models during the helium shell flashes in AGB stars.
It is shown that the distribution of the {\it s}-process elements is independent of the initial metallicity of [Fe/H] $\lesssim -2$.
We also show that the convective $^{13}$C burning is the most important contributor to the {\it s}-process at low-metallicity.
These results imply that the origin of CEMP-no stars can be explained by the binary scenario.

We have demonstrated that the {\it s}-process is contributed by two modes: convective $^{13}$C burning and $^{22}$Ne burning.
It would be crucial to determine the parameters of hydrogen mixing and binary parameters, and their dependences on metallicity, to give an insight into the star and binary formation of the early Galaxy.
More detailed discussion will be given in a separate paper \cite{Yamada2016}.

\section*{Acknowledgement}

This work has been supported by JSPS KAKENHI Grant Number 16H02168.

\end{document}